# Highly responsive UV-photodetectors based on single electrospun $TiO_2$ nanofibres

Aday J. Molina-Mendoza,[a]* Alicia Moya,[b] Riccardo Frisenda,[c] Simon A. Svatek,[c] Patricia Gant,[c] Sergio Gonzalez-Abad,[c] Elisa Antolin,[d] Nicolás Agraït,[a,c,e] Gabino Rubio-Bollinger,[a,e] David Perez de Lara,[c] Juan J. Vilatela,[b*] and Andres Castellanos-Gomez.[c*]

[a] Departamento de Física de la Materia Condensada, Universidad Autónoma de Madrid, Campus de Cantoblanco, E-28049, Madrid, Spain. E-mail: aday.molina@uam.es

[b] IMDEA materials institute, Eric Kandel 2, Getafe, Madrid, 28906 Spain. E-mail: juanjose.vilatela@imdea.org

[c] Instituto Madrileño de Estudios Avanzados en Nanociencia (IMDEA-nanociencia), Campus de Cantoblanco, E-28049 Madrid, Spain. E-mail: andres.castellanos@imdea.org

[d] Instituto de Energía Solar – Universidad Politécnica de Madrid ETSI Telecomunicación, Ciudad Universitaria sn, 28040 Madrid, Spain.

[e] Condensed Matter Physics Center (IFIMAC), Universidad Autónoma de Madrid, E-28049 Madrid, Spain.

## Abstract

In this work we study the optoelectronic properties of individual $TiO_2$ fibres produced through coupled sol-gel and electrospinning, by depositing them onto pre-patterned Ti/Au electrodes on $SiO_2$/Si substrates. Transport measurements in the dark give a conductivity above $2 \cdot 10^{-5}$ S, which increases up to $8 \cdot 10^{-5}$ S in vacuum. Photocurrent measurements under UV-irradiation show high sensitivity (responsivity of 90 $A \cdot W^{-1}$ for 375 nm wavelength) and a response time to illumination of ~ 5 s, which is superior to state-of-the-art $TiO_2$-based UV photodetectors. Both responsivity and response speed are higher in air than in vacuum, due to oxygen adsorbed on the $TiO_2$ surface which traps photoexcited free electrons in the conduction band, thus reducing the recombination processes. The photodetectors are sensitive to light polarization, with an anisotropy ratio of 12%. These results highlight the interesting combination of large surface area and low 1D transport resistance in electrospun $TiO_2$ fibres. The simplicity of the sol-gel/electrospinning synthesis method, combined with a fast response and high responsivity makes them attractive candidates for UV-photodetection in ambient conditions. We anticipate their high (photo) conductance is also relevant for photocatalysis and dye-sensitized solar cells.

## Introduction

Transition metal oxides offer great opportunities in optoelectronic applications requiring transparent materials (materials that scarcely react to light within the visible part of the spectrum) with high responsivity to UV light. UV photodetectors have gained attention due to the wide spectrum of possible applications in fields such as flame sensing, missile plume detection, medical diagnosis, chemical analysis or optical communications.[1, 2] Among the different transparent materials, $TiO_2$ is one of the most studied partially because of its interesting properties and promising applications in photocatalysis, solar cells, batteries or gas sensors.[3-14] Moreover, from the synthesis point of view, $TiO_2$ results especially relevant because of the different synthetic methods that can be employed to obtain this material. For instance, $TiO_2$ UV photodetectors fabricated by sol-gel, potentiostatic anodization or radio-frequency magnetron sputtering have been recently demonstrated.[9, 10, 15] The best photodetector performances reported for a wavelength of 375 nm are close to ~ 25 A/W, with response times in the order of 6 s to 15 s.[9] In this work we present UV photodetector devices based on individual electrospun $TiO_2$ nanofibres transferred onto pre-patterned electrodes. The electrospinning technique becomes highly interesting in this kind of applications since it enables the synthesis of large scale material, simplifying the fabrication of functional devices. The fabricated devices show an outstanding UV photoresponse of ~ 90 A/W and response time of a few seconds, which is above state-of-the-art for $TiO_2$. The photodetectors also show sensitivity to the incident light polarisation, with a polarisation anisotropy of 12%.

## Results and discussion

The $TiO_2$ nanofibres are produced by a combination of electrospinning and sol-gel as reported elsewhere.[16] Briefly, a solution of sol-gel precursors and PVP is electrospun as continuous thin nanofibres. Subsequent pyrolysis and annealing in inert gas removes the polymer coating and completes the crystallisation of $TiO_2$, resulting in a vacancy-rich oxide with high activity in photocatalytic hydrogen production[16] and $CO_2$ reduction under UV irradiation.[17]

Fig. 1a shows a scanning electron microscopy (SEM) image of an array of electrospun $TiO_2$ nanofibres with average diameter of 200 ± 100 nm and only less than 3% exceeding 500 nm. Each fibre consists of a network of $TiO_2$ nanocrystals with average size of 12 nm that form a high-surface area (~ 40$m^2$/g) mesoporous continuous structure (Fig. 1b). High magnification transmission electron micrographs (HRTEM) show the tight interfaces between adjacent nanocrystals, resembling grain boundaries (Fig. 1c). The interconnection of crystalline domains in a continuum, a consequence of the synthetic route used, provides a conduction path with low resistance, particularly compared to nanoparticles simply aggregated by weak interactions.

X-Ray Diffraction (XRD) and Raman spectroscopy with a 532 nm-laser show predominance of the anatase phase (Fig. 1d and e). A small fraction of rutile is also detected, which forms due to the accelerated anatase-rutile phase transformation at the interface between nanocrystals,[16] and is thus to some extent inevitable in these samples. The average anatase crystal size calculated from XRD is 11.22 nm, which matches TEM observations. UV-Vis diffuse reflectance spectra shows an absorption edge below 400 nm, confirming that the material absorbs only in the UV region and can thus be considered transparent (Fig. 1f). The calculated bandgap following the Kubelka-Munk theory is 3.11 eV (398 nm), which is between those for bulk anatase and rutile.[18]

A single $TiO_2$ nanofibre can be isolated to fabricate a device by means of a recently developed "pick-up and drop" technique to pick up a fibre from the substrate and to deterministically place it onto a desired acceptor substrate. In order to do so, we use a transfer setup originally designed for deterministic placement of 2D-materials (see Supporting Information of Ref.[[19]] for details on the setup). We direct the reader to the Supporting Information to see pictures acquired at different stages of the pick-up and drop process used to fabricate a $TiO_2$ fibre-based photodetector (Fig. S1 of Supporting Information). Fig. 2a shows an SEM image of a device fabricated by placing an individual $TiO_2$ fibre between two pre-patterned Ti/Au electrodes. The inset in Fig. 2a shows a high resolution SEM image that enables accurate determination of the diameter of the specific fibre analysed, in this case $d$ = 537.71 nm.

In order to characterize the optoelectronic performance of the device, the light of a high power LED source ($\lambda$ = 455 nm) is focused down to a 200 μm-diameter spot onto the sample. The power is measured with a silicon photodetector (Thorlabs power meter PM100D with sensor S120VC). Fig. 2b shows a comparison of current-voltage (IV hereafter) characteristics acquired in dark conditions and upon illumination with increasing light power ($P$). Interestingly, even in dark conditions the material is significantly conductive. At 10V, for example, it has a conductivity of ~ $6\cdot10^{-4}$ $S\cdot m^{-1}$, which is comparable to that of  monolithic $TiO_2$ with grain size in the range of tens of microns,[20] but orders of magnitude superior to that of sintered mesoporous $TiO_2$ materials, ordered mesoporous $TiO_2$ from block-copolymer directed growth, as well as other morphologies used in photodetectors, all in the $10^{-8}$ S/m range.[9, 10, 21, 22] This is due to the network of nanocrystals discussed above, which minimises the activation energy for charge transfer between adjacent crystalline domains, with the added benefit of oxygen vacancies acting as donors. Note that whereas the majority of mesoporous $TiO_2$ is produced by sintering of pre-synthesised particles, our synthetic method forces interconnection of crystalline domains at the point of nucleation and growth during sol-gel, leading to tight interfaces with large contact area per particle. Additionally, we point to observations of a high density of sub-band gap states near the conduction band in ordered mesoporous $TiO_2$ grown via block-copolymer self-assembly and which lead to large enhancements in dye-sensitised solar cell performance.[23] These states are attributed to the formation of oxygen vacancies as a consequence of the reducing atmosphere during growth in the vicinity of the polymer phase, similar to those observed in our electrospun material, for example in Fig. 1f.

The upper inset of Fig. 2b shows the absolute value of the IV characteristics in logarithmic scale to facilitate the comparison between the different light powers. One can see how the ratio between the dark current and the current upon illumination can reach up to ~ 200 nA with an applied voltage of 10 V. The lower inset shows the photocurrent ($I_{ph}$, difference between the current upon illumination and in dark conditions) as a function of the light spot power. For purely photoconductance photogeneration mechanism (where each photon generates an electron-hole pair that is separated by the applied voltage bias thus increasing the effective conductance of the device) one would expect a linear $I_{ph}$ vs. $P$ trend. Our data, however, shows a sublinear trend that can be fitted to a power law $I_{ph}$ vs. $P^{\alpha}$ with $\alpha$ = 0.72. It has been shown how such a behaviour is observed for systems with a strong photogating effect, where the electrons (or holes) get immobilized in charge traps. The electric field generated by these charged impurities effectively dopes the material by electric field-effect, increasing the conductance of the sample. Photogating effect typically results in a quantum yield higher than 1 but slow responsivities as the devices cannot response faster than the lifetime of the charged traps.

The photoresponse of the device at different wavelengths has been explored by using LED sources with different central wavelengths while fixing the incident power. Fig. 2c shows the photocurrent generated upon illumination with light with different wavelengths (10 V, 15 $W \cdot m^{-2}$). The sharp increase at $\lambda = 375$ nm matches very well with the absorption data shown in Fig. 1f for a thin film of multiple fibres. Along the visible part of the spectrum, the response of the device is negligible compared to that in the UV, demonstrating its potential for applications requiring transparent materials. In order to facilitate the comparison between different photodetectors, it is common to use the responsivity, defined as $R = I_{ph}/P_{eff}$, where $P_{eff}$ is the effective power of light reaching the device and is calculated as $P_{eff} = P_{laser} \cdot A_{dev} / A_{spot}$, ($A_{dev}$ is the surface of the device that "sees" the light and $A_{spot}$ is the total area of the LED spot reaching the device). The inset in Fig. 2c shows the responsivity as a function of the LED wavelength in logarithmic scale to facilitate the comparison between the UV and the VIS part of the spectrum.

The response time of our $TiO_2$ device is characterized by modulating the intensity of the LED source with a square signal (frequency 100 mHz). Fig. 3 shows the photocurrent vs. time measured with different maximum illumination power. The response rise time of a photodetector is defined as the time difference between 10% and 90% of the maximum photocurrent, the fall time is defined in the opposite way. From Fig. 3 we obtain a rise time of 2.5 s and a fall time of 10 s. This relatively slow response time points again towards photogating as the main generation mechanism. Interestingly, the device also shows a fast response occurring in the first 100 ms after the illumination is turned on/off. Therefore, it seems that the most likely scenario is a combination of different photocurrent generation mechanisms where photogating plays a major role. The photogating effect is an especial case of the photoconductive effect in which one type of the photogenerated charge carriers (electrons or holes) gets trapped in localized energy states created by defects or at the surface of the semiconductor. The photogating effect can be distinguished from the photoconductive effect by their response times, which are much slower in the first case (photogating), although a combination of the two of them is also possible.[24, 25] In our case, the photoconductive effect is probably triggering the photoresponse in the first few 100 milliseconds while the photogating effect is slowing down the global response. We direct the reader to Figs. S2, S3, S4 and S5 in the Supporting Information for the optoelectronic characterization of other examples of $TiO_2$ nanofibre photodetectors showing the reproducibility of the results discussed above.

Comparison with some state-of-the-art UV-photodetectors (table 1) yields a very good performance of our $TiO_2$ nanofibres-based photodetectors, which show high responsivity and short response time for a wavelength close to the absorption limit of $TiO_2$ ($\lambda = 375$ nm). Therefore, electrospun $TiO_2$ photodetectors present themselves as great candidates for UV-photodetection with high sensitivity and fast response.

To further understand the photogeneration mechanisms working in our $TiO_2$ photodetectors, we perform the optoelectronic characterization of the same photodetector in air and in vacuum. In dark conditions and vacuum ($P = 7{\cdot}10^{-6}$ mbar) the device has significantly higher conductance (~ 50%) than in air (Fig. 4a and Fig. S8). This feature has been observed in several metal oxide materials such as $SnO_2$ or ZnO,[26, 27] and it is generally attributed to the presence of oxygen molecules adsorbed on the $TiO_2$ surface that trap free electrons from the conduction band ($O_2\,(g) + e^{-} \rightarrow O_2^{-}\,(ad)$) forming a low-conductivity depletion layer near the surface,[28, 29] resulting in the reduction of the channel conduction and thus a reduction of the conduction of the material surface.[10, 30] When the photodetector is illuminated, we see that the time response of the material is strongly dependent on the atmospheric conditions: in vacuum the rise/fall times have values of 23 s / 185 s, while in air is much faster (1.5 s / 7.8 s). This feature has been reproduced in more devices measured in different atmospheres (Fig S9).The slow time response in vacuum, compared to that in oxygen/air, has also been observed in previous works,[10, 26, 27] and is generally attributed to the suppression of oxygen readsorbtion in vacuum, although we also observe that the response in air is noticeably faster than that in oxygen, the adsorbed water or nitrogen molecules might be also affecting the time response of the material. The combination of a high surface area (40m$^2$/g) meosoporous structure and a large fraction of O vacancies implies that surface defects and their interaction with adsorbed molecules play a dominant role in the photoconduction mechanism in electrospun $TiO_2$ fibres.

We have also performed the optoelectronic characterization of another photodetector based on a $TiO_2$ nanofibre crystallised in air atmosphere instead of Ar (Fig. S6 and S7). This sample had an slightly lower photocurrent, although the calculated responsivity is similar to the one obtained for Ar-crystallised $TiO_2$. Further investigation is needed to go deeper in the role of different defects and adsorbates in the photoresponse of these devices, but it lays beyond the scope of the current manuscript.

Another interesting aspect of light absorption in photodetectors based on elongated semiconducting nanomaterials, such as nanowires or nanofibres, is that they usually exhibit polarisation sensitivity to the exciting light.[31-33] We have studied the polarisation sensitivity in $TiO_2$ nanofibres-based photodetectors by varying the polarisation of the incident light. In Fig. 5a we show an artistic drawing of the experimental setup used to measure the polarisation sensitivity in our photodetector: a linear polariser is placed in between the zoom lens (used to illuminate the sample with unpolarised light provided by the LED source) and the sample, making the light reaching the sample to be linearly polarised within the sample plane (Fig. 5a). The polariser is then rotated, with a constant angular frequency (0.07 rad/s) while the photocurrent was recorded, obtaining the photocurrent curve shown in Fig. 5b and c, where we see that the photocurrent is higher when the light is linearly polarised parallel to the nanofibre longitudinal direction and lower when it is perpendicular. We extract the polarisation anisotropy, defined as $\rho = (I_{\parallel} - I_{\perp}) / (I_{\parallel} + I_{\perp})$, where $I_{\parallel}$ and $I_{\perp}$ are the photocurrent with incident light polarised parallel or perpendicular to the main axis of the fibre, respectively, obtaining $\rho = 12\% \pm 2\%$. In a nanowire with a radius comparable to the incident light wavelength, the optical absorption is dramatically determined by the polarization of the incident light, being higher for light with polarization parallel to the nanowire than for perpendicular polarization,[34] since photocurrent linearly depends on the absorption, it will also be polarization-dependent. This emphasises the 1-dimensional character of electrospun $TiO_2$ and suggests its potential use as polarisation-sensitive photodetector.

## Conclusions

In summary, we have studied the photoresponse of single electrospun $TiO_2$ nanofibre-based photodetectors in a wide range of the electromagnetic spectrum (from 375 nm to 1050 nm), finding a good performance in the UV with responsivity values up to 90 $A \cdot W^{-1}$ with an applied voltage of 10 V with 375 nm wavelength and a power density of 15 $W \cdot m^{-2}$. The photodetectors show a time response to the incident light of ~ 5 s, as well as polarisation sensitivity with an anisotropy ratio of 12%. Dark current measurements show $TiO_2$ nanofibres to have a very high conductivity, reminiscent of the synthetic process used and which leads to the formation of a 1-dimensional mesoporous network of interconnected crystalline domains. Work is in progress to determine the relative contributions of low internal resistance and vacancies (consequence of the Ar annealing) to their transport properties. Thus, the fibres have relatively large surface area and are sensitive to adsorbed $O_2$, which acts as electron scavenger and reduces conductivity compared to vacuum. However, adsorbed $O_2^-$ ions trap photogenerated holes and thus, both photocurrent and response speed are higher in air. The excellent performance of our nanofibre $TiO_2$ devices above state-of-the art $TiO_2$-based photodetectors, combined with the simplicity of the synthesis method based on sol-gel and electrospinning, makes these nanofibres strong candidates for UV light detection transparent to VIS light.

## Acknowledgements

A.C-G. acknowledges financial support from the BBVA Foundation through the fellowship "I Convocatoria de Ayudas Fundacion BBVA a Investigadores, Innovadores y Creadores Culturales", from the MINECO (Ramón y Cajal 2014 program, RYC-2014-01406), from the MICINN (MAT2014-58399-JIN) and the European Commission under the Graphene Flagship, contract CNECTICT-604391. A.J.M-M., G.R-B., S.A.S and N.A. acknowledge the support of the MICINN/MINECO (Spain) through the programs MAT2014-57915-R, BES-2012-057346 and FIS2011-23488, Comunidad de Madrid (Spain) through the programs NANOBIOMAGNET (s2009/MAT-1726) and S2013/MIT-3007 (MAD2D) and the European Commission through the FP7 ITN MOLESCO (Project Number 606728). A.M. and J.J.V. acknowledge the support of European Union Seventh Framework Program under grant agreements 310184 (CARINHYPH project), MINECO (MAT2015-62584-ERC, RyC-2014-15115, Spain) and the Madrid regional government (S2013/MIT-3007 MAD2D project). D.PdL. acknowledges the support of MICINN/MINECO (Spain) through the program FIS2015-67367-C2-1-P. R.F. acknowledges support from the Netherlands Organisation for Scientific Research (NWO) through the research program Rubicon with project number 680-50-1515.

## Figures

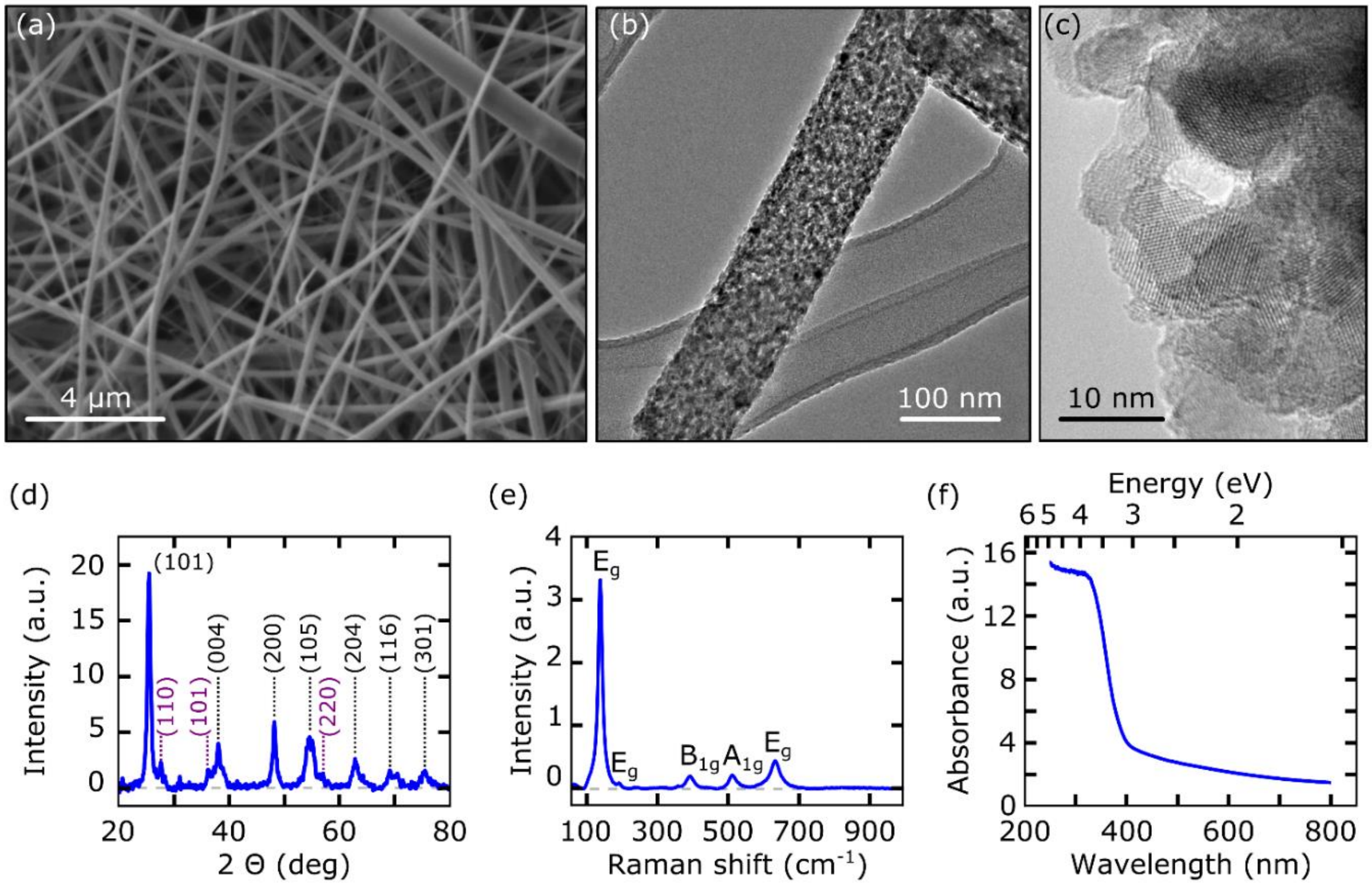

**Fig. 1 (a-c)** Scanning and Transmission electron micrographs of the mesopororous fibre structure formed by interconnection of $TiO_2$ nanocrystals. **(d)** XRD pattern (the black lines indicate the anatase reflexions and the purple lines indicate the rutile reflexions) and **(e)** Raman spectra of $TiO_2$ nanofibres showing predominantly anatase phase (anatase vibrational modes are highlighted in the figure). **(f)** UV-Vis diffuse reflectance spectra shows absorption of $TiO_2$ nanofibres in the UV range, below 400 nm.

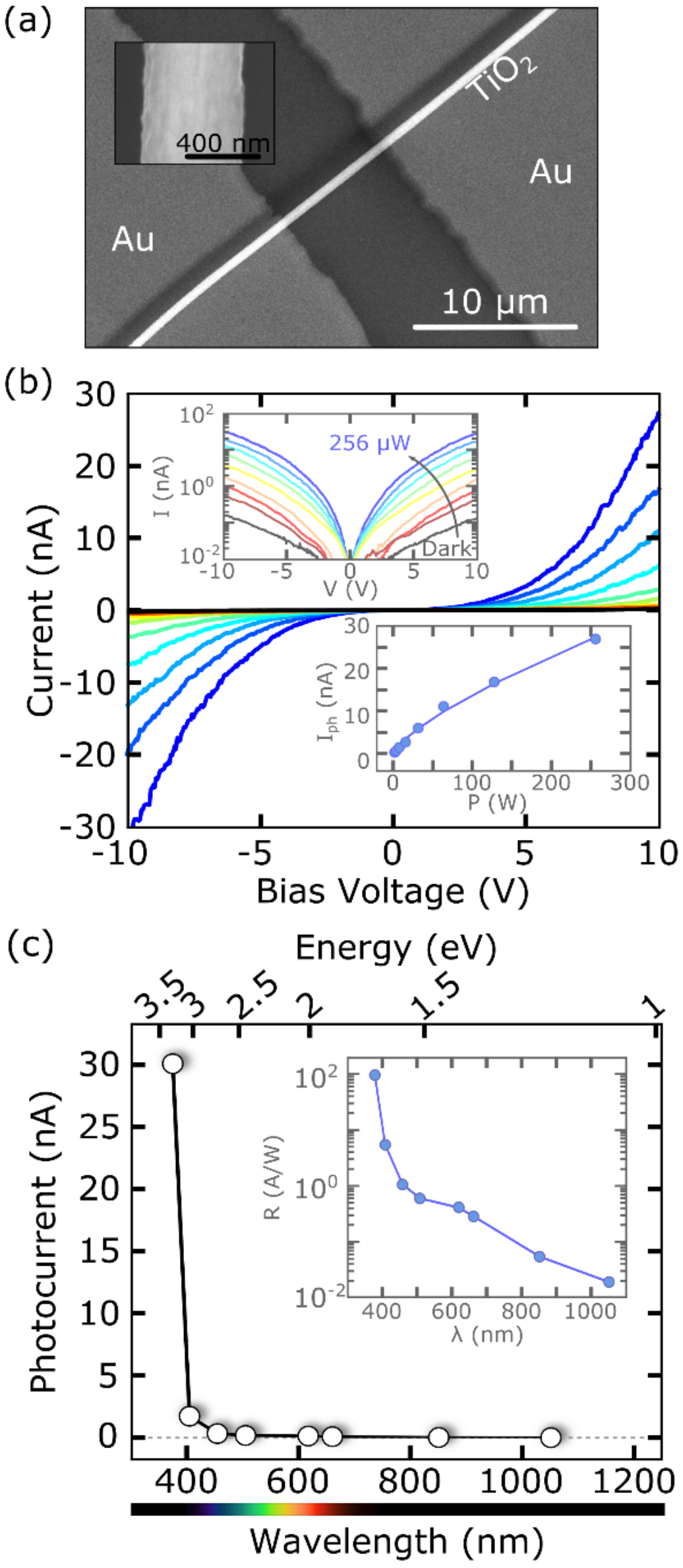

**Fig. 2 (a)** SEM image of a $TiO_2$ fibre photodetector. Inset: zoom of the fibre. **(b)** Current-voltage characteristics of the $TiO_2$ photodetector shown in (a) in dark conditions and upon illumination with 455 nm wavelength with increasing LED power. Upper inset: the same current-voltage curves in logarithmic scale. Lower inset: photocurrent as a function of the LED power. **(c)** Photocurrent of the device shown in (a) as a function of the LED wavelength ($P$ = 2 μW, $V_b$ = 10 V). Inset: responsivity as a function of the LED wavelength.

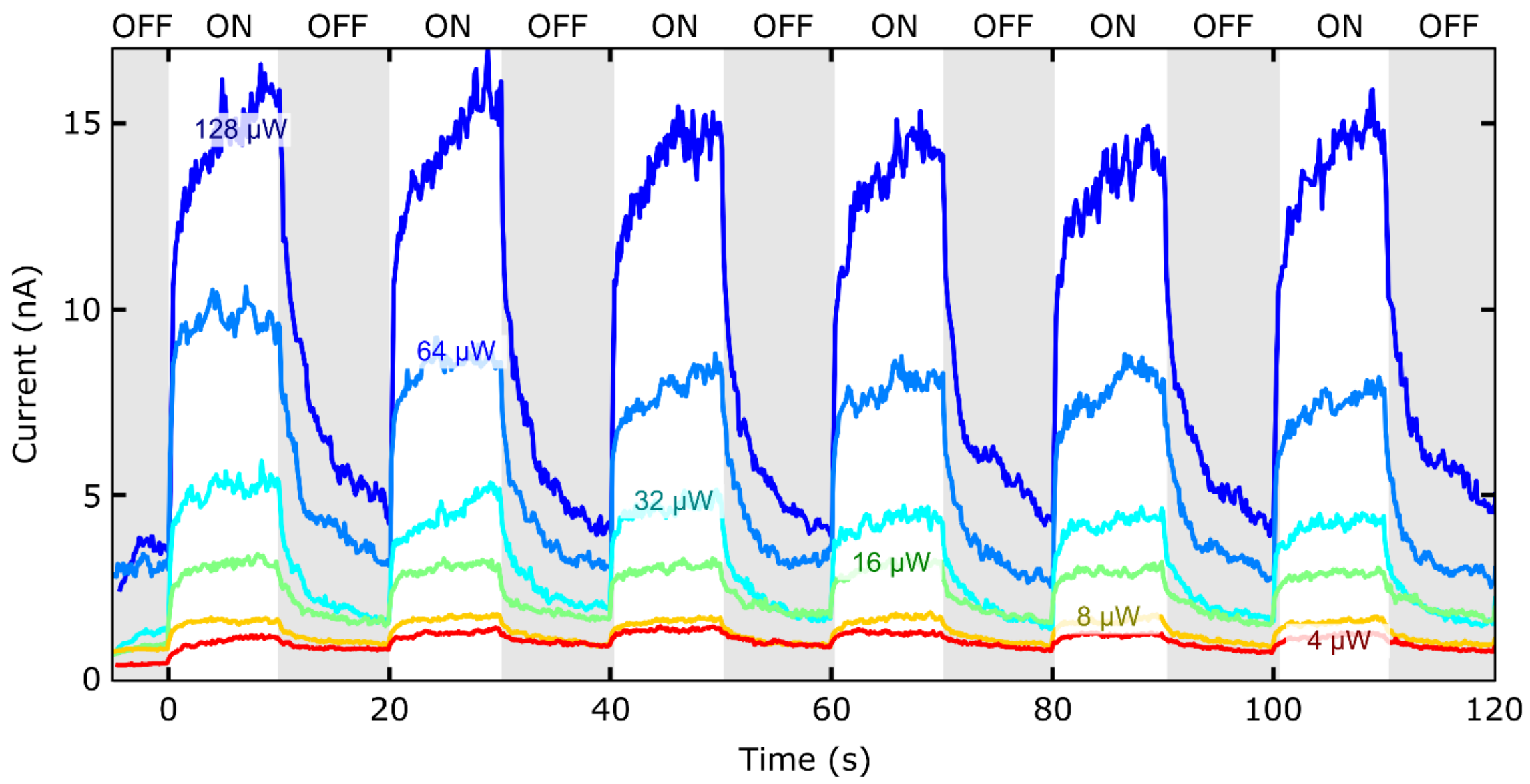

**Fig. 3** Time response of the photodetector shown in Fig. 2 upon illumination with 455 nm wavelength with increasing LED power. In order to highlight the photocurrent, the dark current has been set to 0. The measured rise time is ~ 2.5 s and the fall time is ~ 10 s.

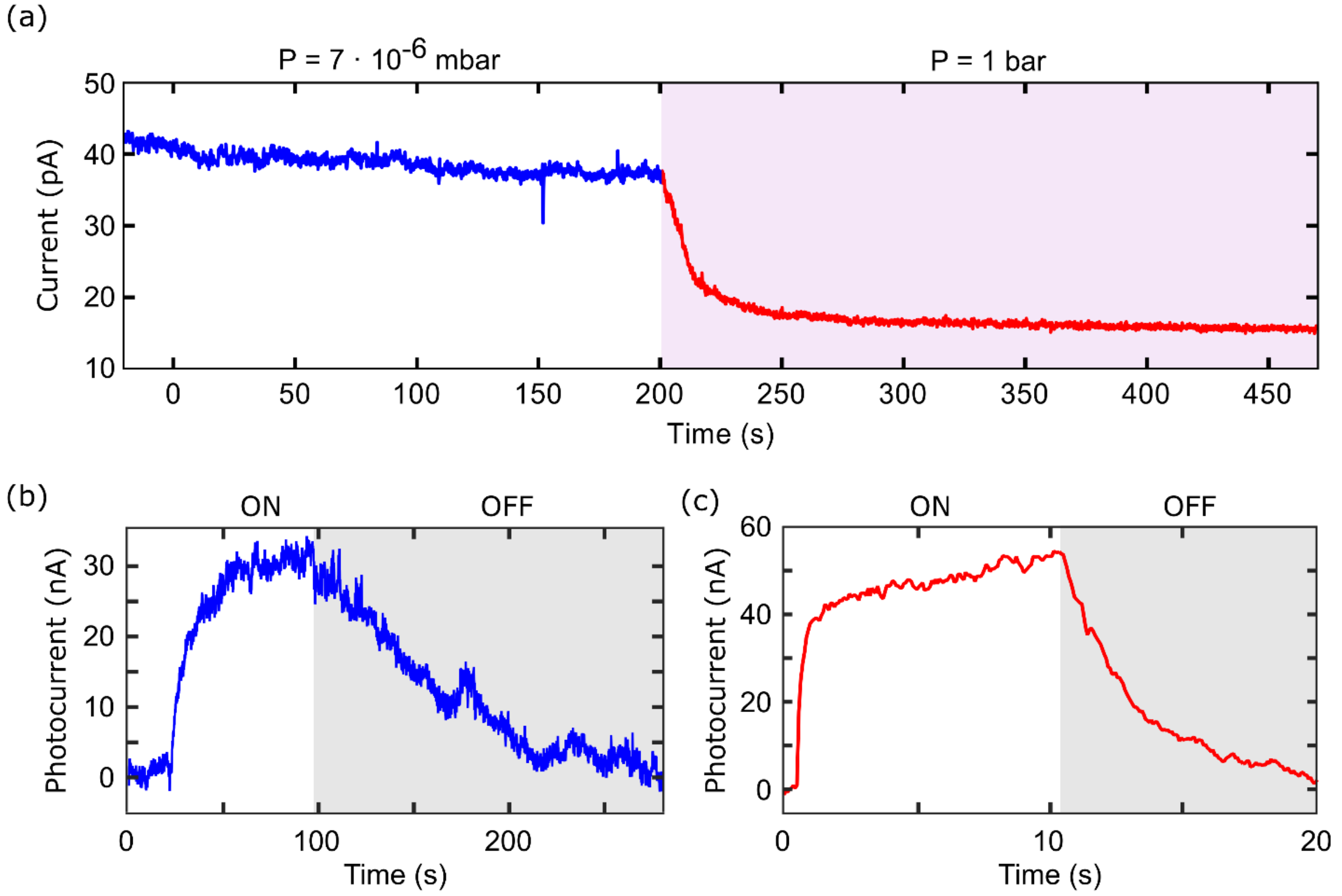

**Fig. 4 (a)** Current as a function of time in a $TiO_2$ nanofibre photodetector in dark conditions both in vacuum (white area) and in air (purple area) with an applied voltage of $V_b$ = 10 V. The current in air is 50% lower than in vacuum. **(b)** Photocurrent of the same device as in (a) in vacuum. **(c)** Photocurrent of the same device as in (a) and (b) in air. The photocurrent in air is 60% higher than in vacuum, and the response time are much faster.

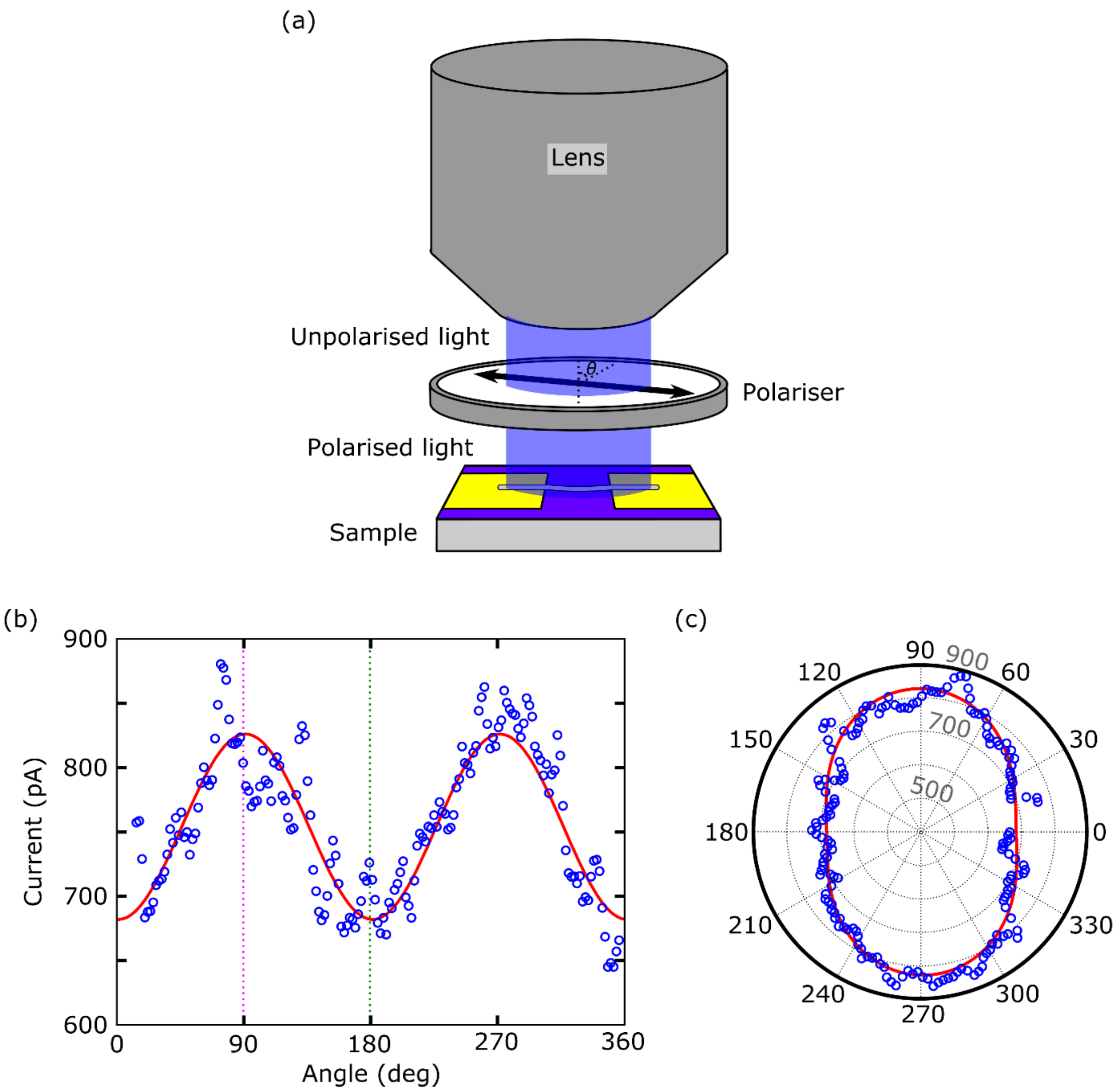

**Fig. 5 (a)** Artistic representation of the experimental setup used to measure the polarisation sensitivity in the $TiO_2$ photodetectors. The light provided by the LED sources is coupled into a zoom lens that creates a light spot on the sample. A linear polariser is placed between the lens and the sample in such a way that the light reaching the device is polarised. The linear polariser is rotated, changing the polarisation angle of the light reaching the device, while the current passing through the device is measured. **(b)** Photogenerated current in a $TiO_2$ nanofibre photodetector as a function of light polarisation (455 nm, 75 $mW \cdot cm^{-2}$) with respect to the longitudinal direction of the fibre, i.e., 0° means that the light is polarised perpendicular to the fibre and 90° means that it is parallel. The blue circles represent the experimental data, while the red line represents a sinusoidal curve superposed to the experimental data. The pink (green) dashed line indicate the angle value of the light polarised parallel (perpendicular) to the nanofibre. **(c)** The same data as in (a) in polar coordinates.

**Table 1.** Nanowire-based UV-photodetectors figures of merit

Thin films

| **Material** | **$V_{ds}$ (V)** | **Wavelength (nm)** | **Power ($W/m^2$)** | **On/Off ratio** | **Responsivity (A/W)** | **Rise time (s)** | **Ref.** |
|---|---|---|---|---|---|---|---|
| ZnO | 1 | UV | $4.5\cdot10^{-3}$ | $1.2\cdot10^{5}$ | - | 360 | [35] |
| $SnO_2$ | 12 | 365 | - | ~ 10 | - | ~ 100 | [36] |
| $TiO_2$ | 10 | 370 | - | - | ~ $4\cdot10^{-2}$ | - | [37] |
| $TiO_2$ Nanotube arrays | 5 | 390 | - | 100 | $7.65\cdot10^{-6}$ | ~ 100 | [38] |

Nanowires

| **Material** | **$V_{ds}$ (V)** | **Wavelength (nm)** | **Power ($W/m^2$)** | **On/Off ratio** | **Responsivity (A/W)** | **Rise time (s)** | **Ref.** |
|---|---|---|---|---|---|---|---|
| ZnO | 1 | 365 | - | $10^{4}$ | - | < 1 | [30] |
| $SnO_2$ | 0.1 | 325 | 100 $W/m^2$ | ~ 10 | 321 | ~ 100 | [39] |
| $SnO_2$ | 1 | 320<br>375* | 9 $W/m^2$ | $10^{3}$<br>- | ~$10^{5}$<br>~ $5\cdot10^{3}$* | ~ 50<br>- | [40] |
| $V_2O_5$ | 1 | 450 | 28 $W/m^2$ | ~ 1.21 | 482 | - | [41] |
| $WO_3$ | 10 | 375 | $1.7\cdot10^{-3}$ | ~ 1.15 | - | ~ 100 | [42] |
| $TiO_2$ | 5 | 250<br>375* | $150\cdot10^{-3}$ | ~ $10^{4}$<br>- | 889.6<br>2* | $13.34\cdot10^{-3}$<br>- | [9] |
| $TiO_2$ | 10 | 375 | 15 | ~ 10 | 90 | 2.5 | This work |

**Table 1.** Comparative table with the figures-of-merit of different UV-photodetectors based on metal oxides in thin film morphology and nanowires. The values marked with * are not explicitly given in the main text of the manuscripts and have been either extracted from plots present in the manuscripts or calculated with the values listed in them.

## Experimental methods

### Synthesis of $TiO_2$ nanofibres

The $TiO_2$ nanofibres are synthetized by a combination of electrospinning and sol-gel methods. A solution of sol-gel precursors (alcoholic solution of titanium ethoxide with controlled pH, 63.5 wt.%) and polyvinyl pyrrolidone (PVP) in ethanol (10 wt.) are mixed with a few drops of acetic acid to catalyse the sol-gel reaction, and then electrospun in a electrospinning setup (Nanon 01A, MECC Co., Ltd.) at 18 kV and flow rate of 2 mL/h to obtain the nanofibres thin film. The polymer coating is removed by pyrolysis in air at 400 °C for 2.5 h and crystallisation was carried out by annealing in Ar atmosphere at 500 °C for 1 h. A reference sample was produced by carrying out the crystallisation process in air. More details of the main features of the $TiO_2$ fibres are presented in the Supporting information.

### Structural and physical characterization of $TiO_2$

The morphology of $TiO_2$ thin films and single nanofibre photodetectors was analysed by scanning electron microscopy (SEM, EVO MA15, Zeiss Model). The structural analysis was carried out by high-resolution transmission electron microscopy (HRTEM, JEOL JEM 3000F). Phase analysis was performed using X-Ray Diffraction (XRD, X´Pert MD Analytical). Micro-Raman spectroscopy was performed with a Renishaw PLC using a laser of 532nm and a power of 5mW. UV-Vis Diffuse Reflectance Spectroscopy was analysed in the range 250-800nm with a Lambda 1050 PerkinElmer spectrometer. Tauc plot representation was used to determine the band gap value of the material.

### Fabrication of $TiO_2$ photodetectors

$TiO_2$ single nanofibre photodetectors are fabricated by "picking up" the nanofibres directly from the thin film: a polydimethylsiloxane (PDMS) stamp (Gelfilm from Gelpak®) is placed on the $TiO_2$ nanofibres thin film and peeled off fast, removing several nanofibres that remain adhered to the stamp. The stamp is then investigated by optical microscopy in order to locate the nanofibres with the desired dimensions (~ 70 µm in length and 200 – 1000 nm in diameter). Since the fibres need to be identified optically, only the fibres with diameters above 400 nm are used. Finally, the nanofibre is transferred deterministically bridging two Ti/Au (5/50 nm) electrodes pre-patterned on a $SiO_2$/Si substrate.

### Optoelectronic characterization of $TiO_2$ photodetectors

The optoelectronic properties of $TiO_2$ photodetectors are characterized in a homebuilt air-pressure (room temperature) probe station. A source-meter source-measure unit (Keithley 2450) is used to perform the current-voltage measurements. The light source is provided by 8 light emitting diodes (LEDD1B – T-Cube LED driver) with different wavelengths ranging from 375 nm to 1050 nm, coupled to a multimode optical fibre at the LED source and directed to the probe station zoom lens, creating a light spot on the sample of 200 µm. The time-dependent measurements are carried out by modulating the light intensity with a function generator (Yokowaga).

# Supporting Information

# Highly responsive UV-photodetectors based on single electrospun $TiO_2$ nanofibres

Aday J. Molina-Mendoza,[a] Alicia Moya,[b] Riccardo Frisenda,[c] Simon A. Svatek,[c] Patricia Gant,[c] Sergio Gonzalez-Abad,[c] Elisa Antolin,[d] Nicolás Agraït,[a,c,e] Gabino Rubio-Bollinger,[a,e] David Perez de Lara,[c] Juan J. Vilatela,[b] and Andres Castellanos-Gomez.[c]

[a.] *Departamento de Física de la Materia Condensada, Universidad Autónoma de Madrid, Campus de Cantoblanco, E-28049, Madrid, Spain. E-mail: aday.molina@uam.es*

[b.] *IMDEA materials institute, Eric Kandel 2, Getafe, Madrid, 28906 Spain. E-mail: juanjose.vilatela@imdea.org*

[c.] *Instituto Madrileño de Estudios Avanzados en Nanociencia (IMDEA-nanociencia), Campus de Cantoblanco, E-28049 Madrid, Spain. E-mail: andres.castellanos@imdea.org*

[d.] *Instituto de Energía Solar – Universidad Politécnica de Madrid ETSI Telecomunicación, Ciudad Universitaria sn, 28040 Madrid, Spain.*

[e.] Condensed Matter Physics Center (IFIMAC), Universidad Autónoma de Madrid, E-28049 Madrid, Spain.

In this supporting information we include the following content:

- Electrospun $TiO_2$ nanofibre synthesis
- Photodetector fabrication method
- More photodetectors characterization
- Photodetectors based on $TiO_2$ nanofibres annealed in air
- Photodetectors performance under different atmospheres

## <u>Electrospun $TiO_2$ nanofibre synthesis</u>

In this work, we use a sol-gel based electrospinning method to produce $TiO_2$ nanofibres. A polymer solution of PVP (polyvinylpyrrolidone, 10%wt.) in ethanol is mixed with an alcoholic solution of metal precursor (titanium ethoxide, 63%wt.) and acetic acid as catalyst of the sol-gel reaction. After reaching a homogeneous solution, it is loaded into a syringe placed in the electrospinning set-up (Nanon 01A, MECC CO., LTD.). A continuous mesh of fibres is collected at 10 cm distance from the tip and using an applied voltage of 18 kV and 2 mL/h of flow rate. After electrospinning, the fibres are thermally treated in air to remove the polymer at 400 °C for 2.5 hours and then annealed in Ar at 500 °C for 1 hour to complete the crystallisation of the metal oxide. In addition, a sample of $TiO_2$ fibre anneled in air at 500°C for 1 hour was also prepared and used for a preliminary evaluation of the effect of the annealing atmosphere on the photoconduction mechanism and properties of devices.

The nanofibre structure consists of a mesoporous network of well-ordered $TiO_2$ nanocrystals and presents two main features different from standard $TiO_2$ produced by sol-gel, namely, 1) the use of the polymer keep the titanium sol unit which accelerates the crystallisation and the anatase-to-rutile phase transformation, creating tight interfaces between the $TiO_2$ nanocrystals and 2) the annealing in inert atmosphere produces oxygen vacancies.[1-3] Both phenomena, interfaces and vacancies, facilitate the creation of internal junction with new electronic states that could trap electrons and ultimately increase lifetime and diffusion coefficients of charge carriers in the material.

## Photodetector fabrication method

In Fig. S1 **1** to **4** we can see in the pictures, through the PDMS stamp, the pre-patterned Au electrodes on a $SiO_2$ substrate and some blurred shapes. These blurred shapes are the $TiO_2$ nanofibres after been peeled off from the thin film, as seen from the backside of the stamp, out of focus. When the stamp is approaching the surface, the flakes get more and more focused and we can align substrate and sample (picture **3**). Finally, when the stamp gets in contact with the substrate, we see a change in the color of both the substrate and the flake (picture **4**). In Fig. S1 **5** to **9** we show pictures of the PDMS been peeled off from the substrate in order to transfer the nanofibers. It is possible to recognize the part of the PDMS which is already peeled off from the one that is still in contact with the substrate by the difference in the color. The line that moves from one picture to the next one is the meniscus that separates the peeled stamp from the part still in contact. Once the meniscus has passed the flake, we can completely remove the stamp.

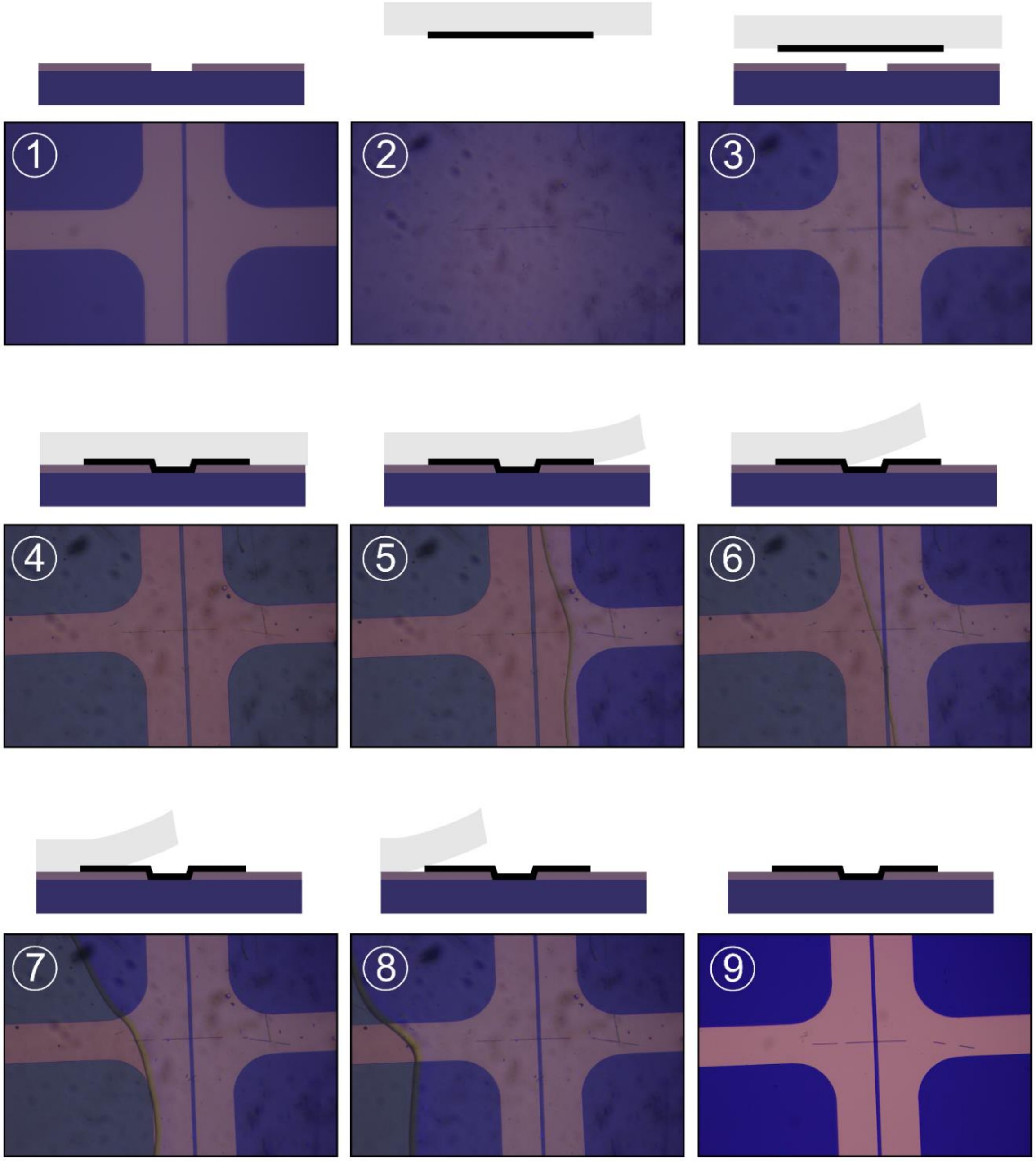

**Fig. S1.** Optical microscopy images of the fabrication of a $TiO_2$ nanofibre-based photodetector. From **1** to **4**: the PDMS stamp is approached to the substrate and the nanofiber is aligned to the electrodes. From **5** to **9**: the PMDS stamp is slowly peeled off to transfer the nanofibre.

## More photodetectors characterization

Photodetector 1

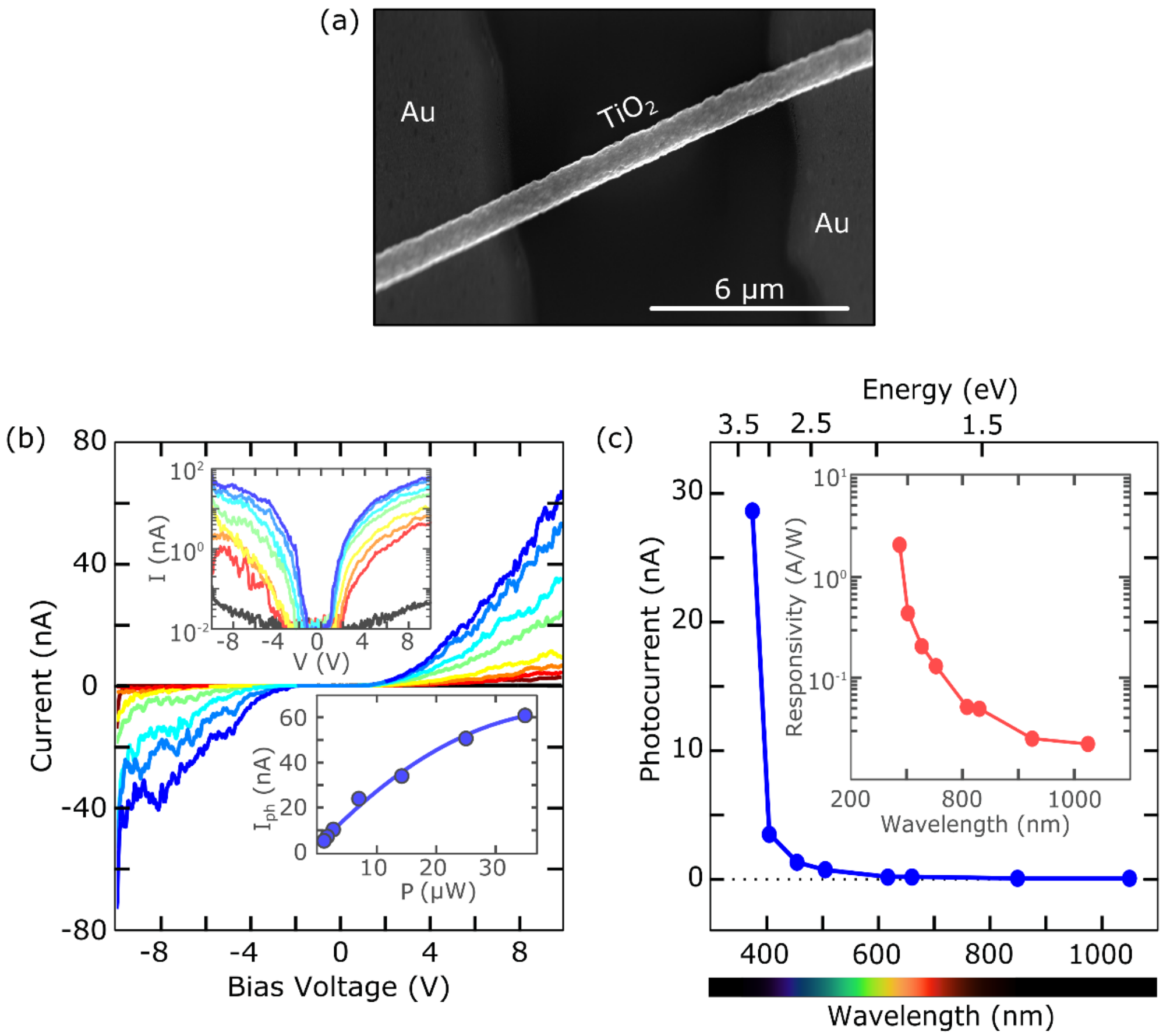

**Fig. S2 (a)** SEM image of a $TiO_2$ photodetector. **(b)** Current-voltage characteristics upon illumination with light wavelength 375 nm excitation for increasing light powers up to 35 µW. Upper inset shows the current-voltage characteristics in a semilogarithmic plot. Lower inset shows photocurrent for increasing powers. The solid line is a power law fit. **(c)** Photocurrent measured as a function of different laser wavelengths ($P$ = 15 µW). Measurements are taken at $V_{ds}$ = 10 V. Inset shows a semilogarithmic plot of the responsivity as a function of the wavelength. Solid lines are guides.

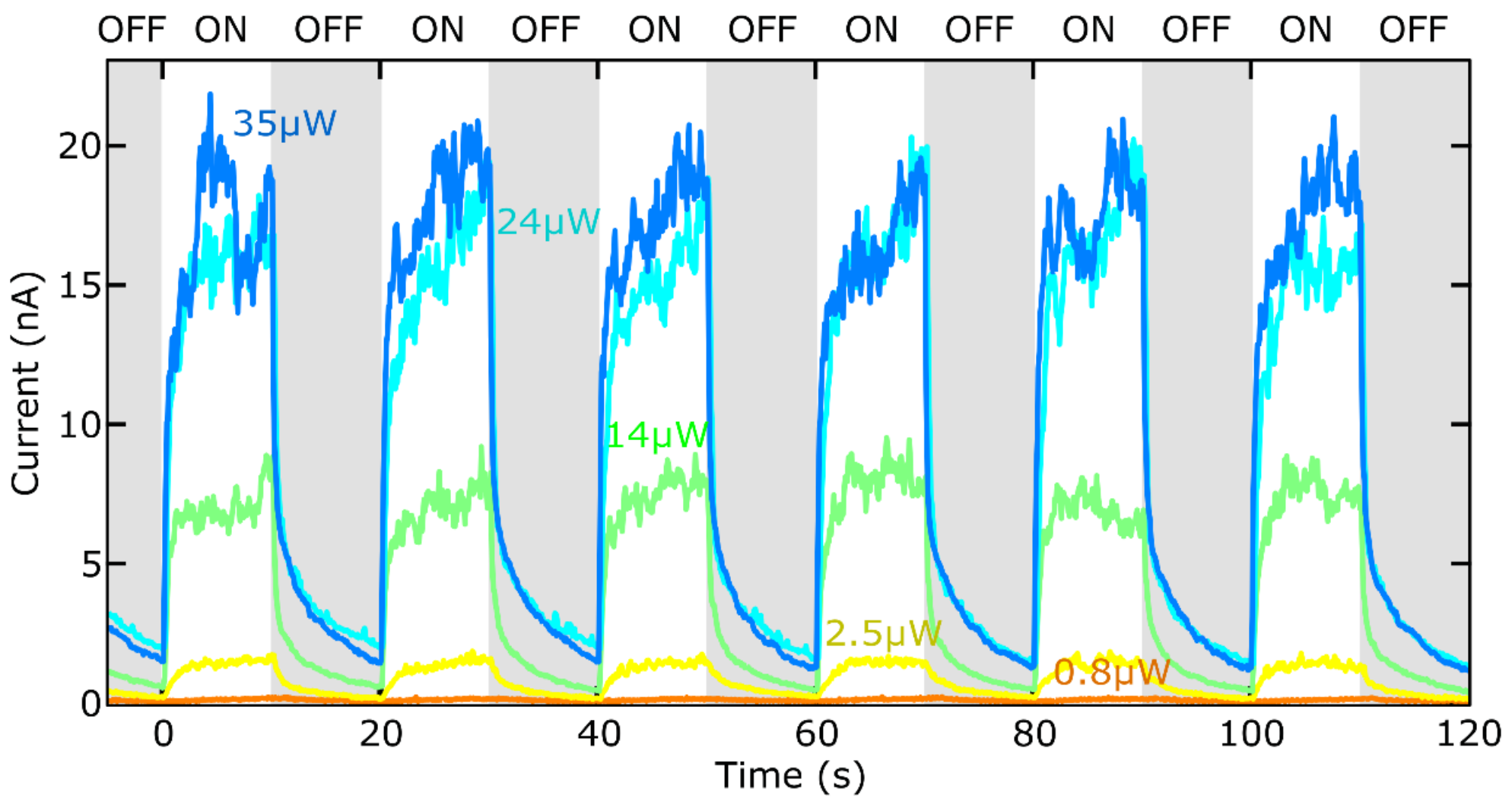

**Fig. S3** Current time response of the device shown in Fig. S2 under a 100 mHz modulated optical excitation (λ = 375nm) for increasing laser powers up to 35 μW. Measurements are acquiered at $V_b$ = 10 V.

Photodetector 2

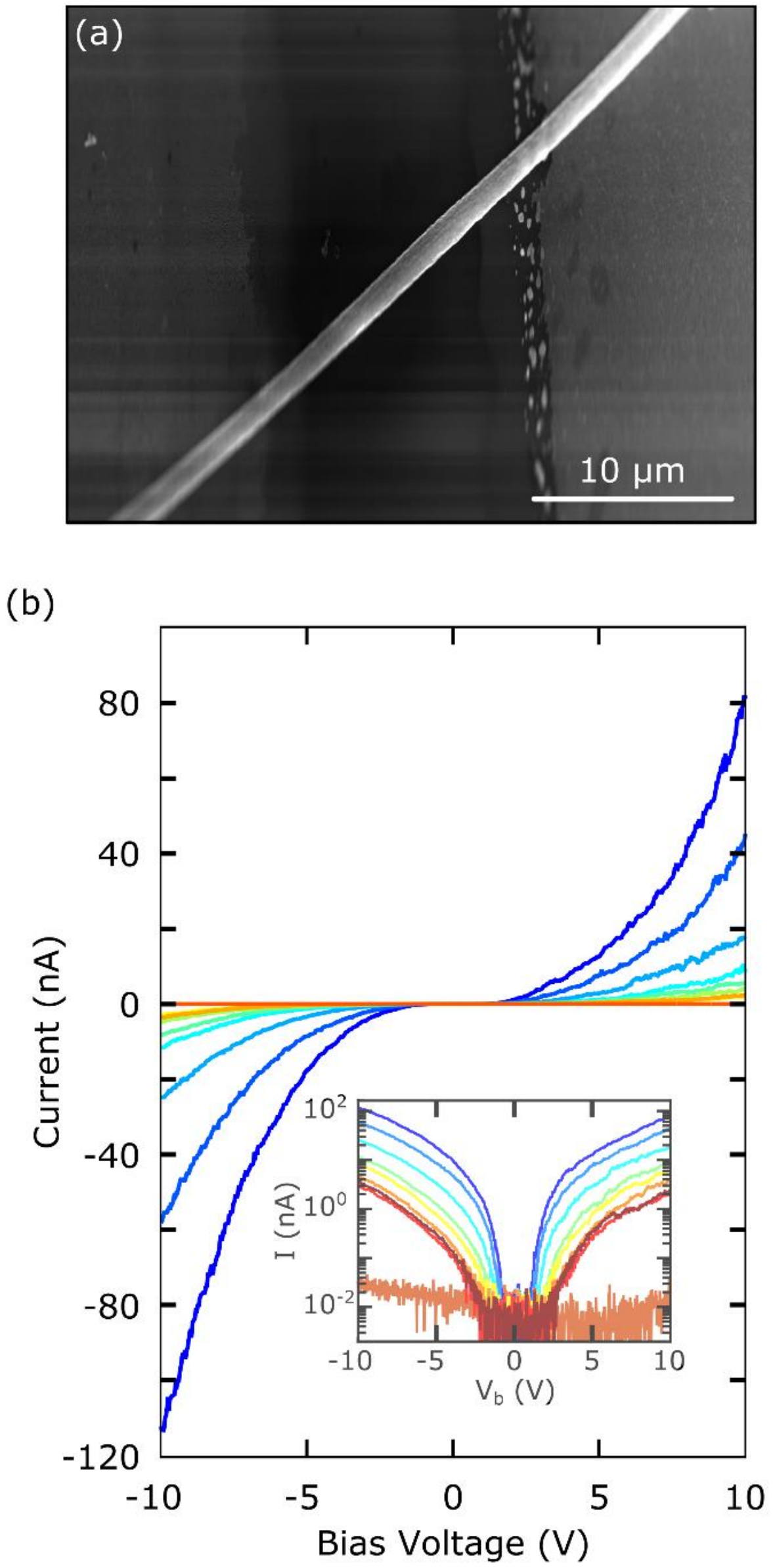

**Fig. S4 (a)** SEM image of a $TiO_2$ photodetector. **(b)** Current-voltage characteristics upon illumination with light wavelength 405 nm excitation for increasing light powers up to 250 μW. The inset shows the current-voltage characteristics in a semilogarithmic plot.

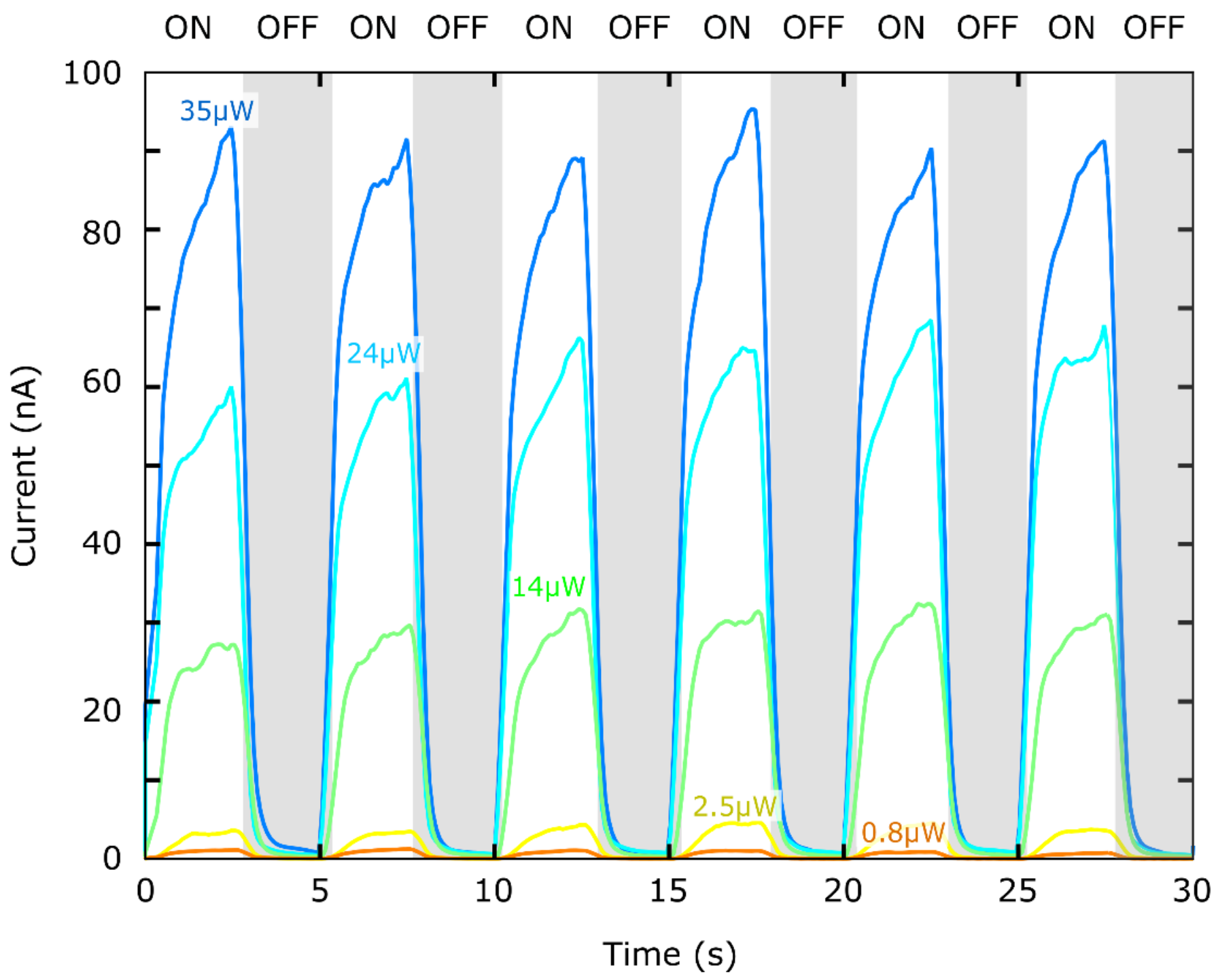

**Fig. S5** Time response of the device shown in Fig. S4 upon a 500 mHz modulated optical excitation (λ = 375nm) for increasing laser powers up to 35 µW. Measurements are taken at $V_b$ = 10 V.

**Photodetectors based on $TiO_2$ nanofibres annealed in air**

We have investigated the role played by the crystallisation atmosphere in the performance of $TiO_2$ nanofibres-based photodetectors. The nanofibre characterized in Fig. S6 and S7 has been synthesized following exactly the same synthetic procedure than the $TiO_2$ fibres from the main text except that it was crystallised in air. The electronic characterization of the device in air (not shown) yields low conductivity with current values lower than 1 pA for $V_b$ = 10 V. We first study the response to incident light of the photodetector by calculating the responsivity as function of the light power (λ = 375 nm, $V_b$ = 10 V, Fig. S6), which reaches a maximum value of ~ 33 A/W.

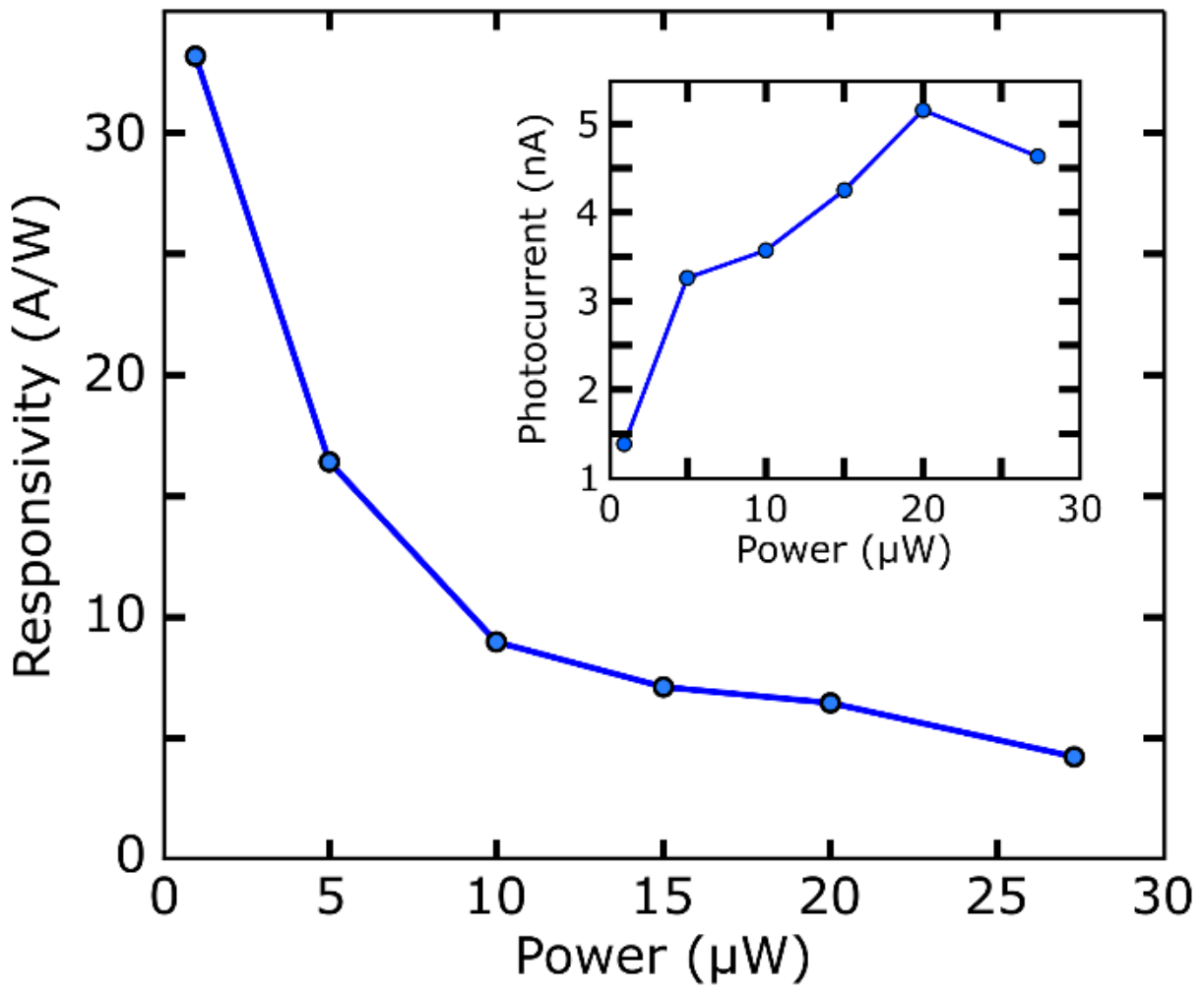

**Fig. S6** Responsivity as a function of the light power of the photodetector based on a single $TiO_2$ nanofibre annealed in air. The maximum value achieved is of 33 A/W. Inset: photocurrent generated in the photodetector as a function of the light power.

We also study the responsivity of the device as a function of the light wavelength, shown in Fig. S7. It is important to note here that the photoresponse as a function of the light wavelength has been measured for the maximum power provided by the LED sources for each wavelength, since the response to lower incident light powers for light wavelengths larger than 405 nm were too small to be measured with our experimental setup. The highest responsivity (~ 4 A/W) in this case is obtained for a light wavelength of λ = 375 nm ($V_b$ = 10 V).

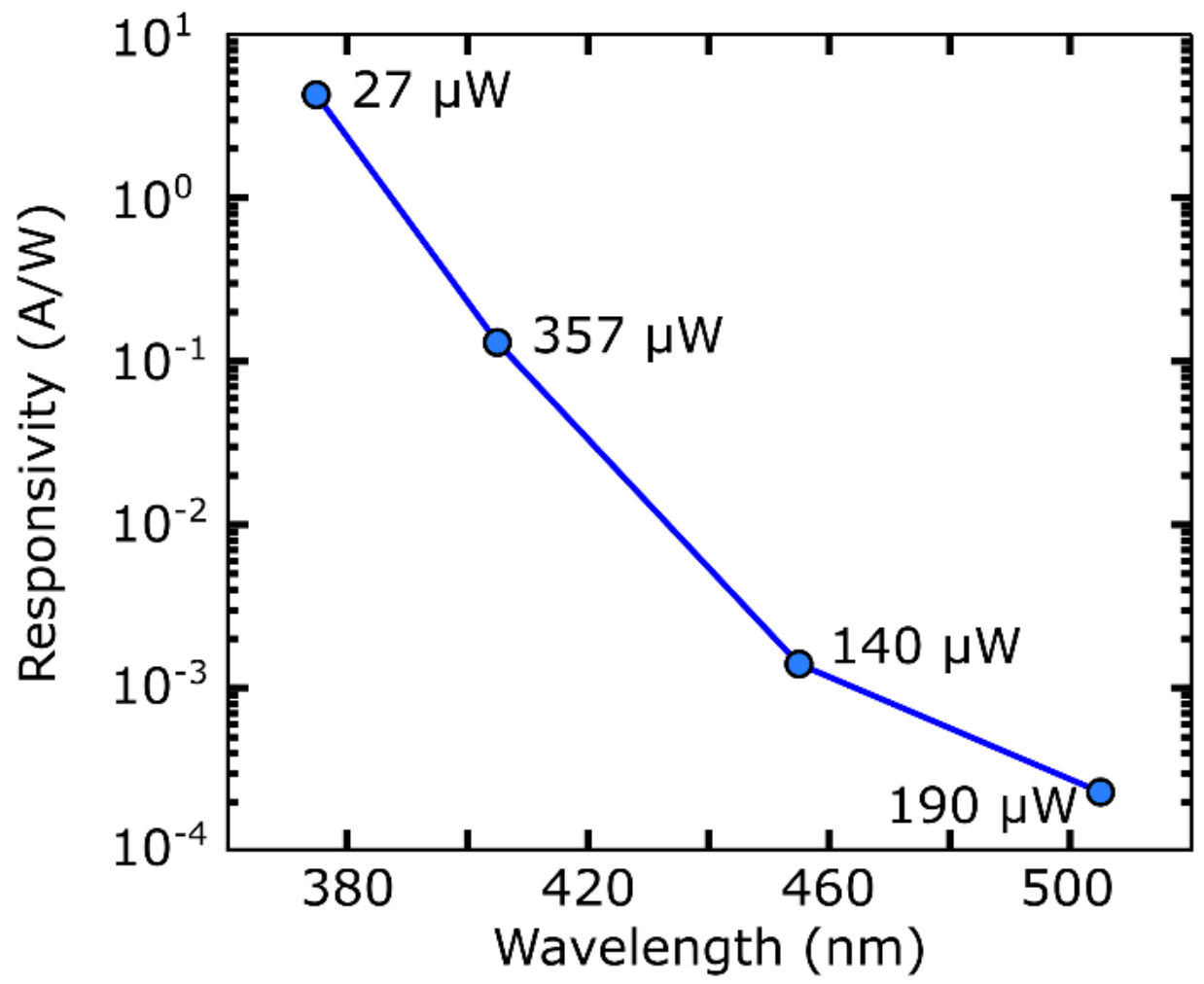

**Fig. S7** Responsivity as a function of the light wavelength of the photodetector based on a single $TiO_2$ nanofibre annealed in air. The maximum value (~ 4 A/W) is obtained for a light wavelength of $\lambda = 375$ nm ($V_b = 10$ V).

**Photodetectors performance under different atmospheres**

The photoresponse of two more devices apart from the one shown in the main text (Fig. 4) have been measured under different atmosphere. The dark current of the devices in vacuum ($10^{-5}$ mbar) and in oxygen atmosphere (1 bar), where it is possible to see that the dark current is enhanced in vacuum with respect to oxygen (Fig. S8a). Another device has been measured in vacuum ($10^{-5}$ mbar) and in air (Fig. S8b), showing a higher dark current in vacuum than in air, giving evidence that the oxygen adsorbed on the nanofibre surface might be playing a role in the conductance in the material.

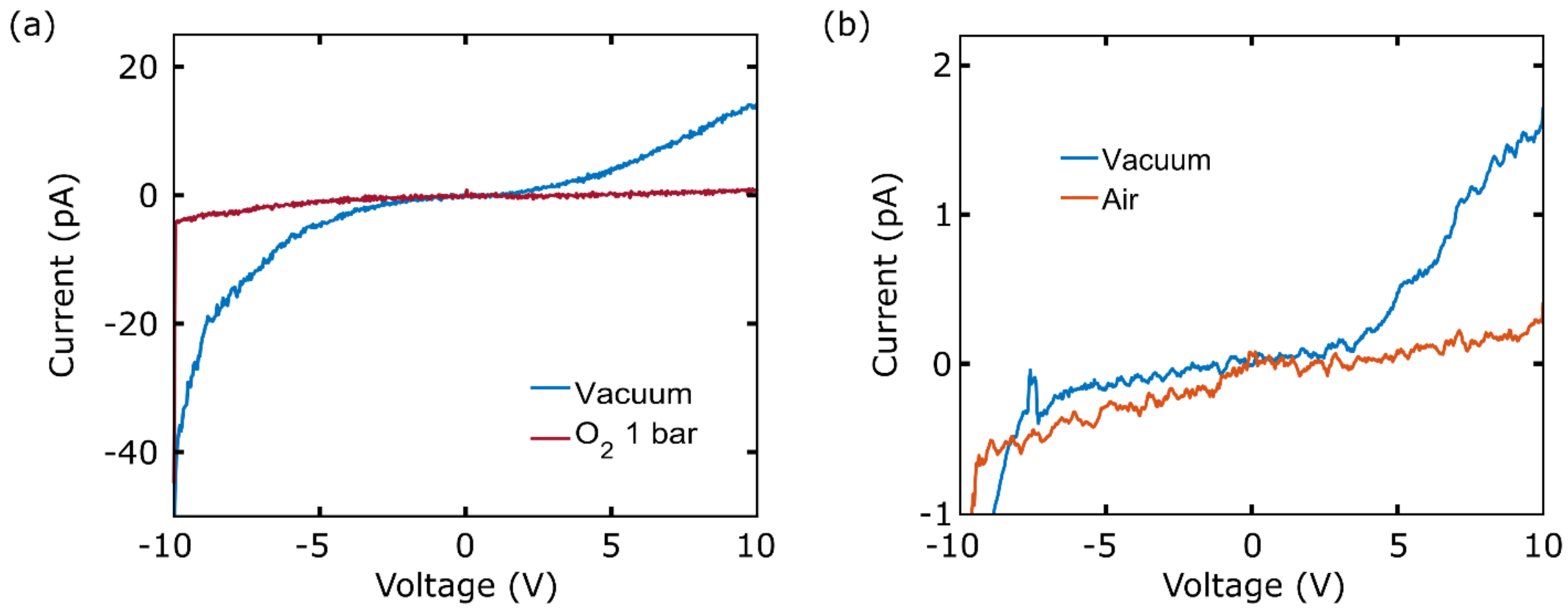

**Fig. S8 (a)** Current-voltage characteristic in dark conditions of a device in vacuum ($10^{-5}$ mbar) and in oxygen (1 bar). The dark current in oxygen is lower than in vacuum. **(b)** Current-voltage characteristic in dark conditions of a device in vacuum ($10^{-5}$ mbar) and in air. The dark current in air is lower than in vacuum.

We have also measured the photoresponse (wavelength of 455 nm, light intensity 0.7 W/cm$^2$) of the devices shown in Fig. S6 as a function of time. The time response in vacuum ($10^{-5}$ mbar) is considerably slower than in oxygen (Fig. S9a) and in air (Fig. S9b), especially regarding the fall time. In Fig. S9a, the fall time of the device in vacuum is >100 s, while in oxygen is ~80 s. In Fig. S9b, the fall time of the device in vacuum is ~90 s, while in air it is <1s, suggesting that the adsorbed oxygen and the water molecules on the fibre surface are playing an important role in the time response of the material.

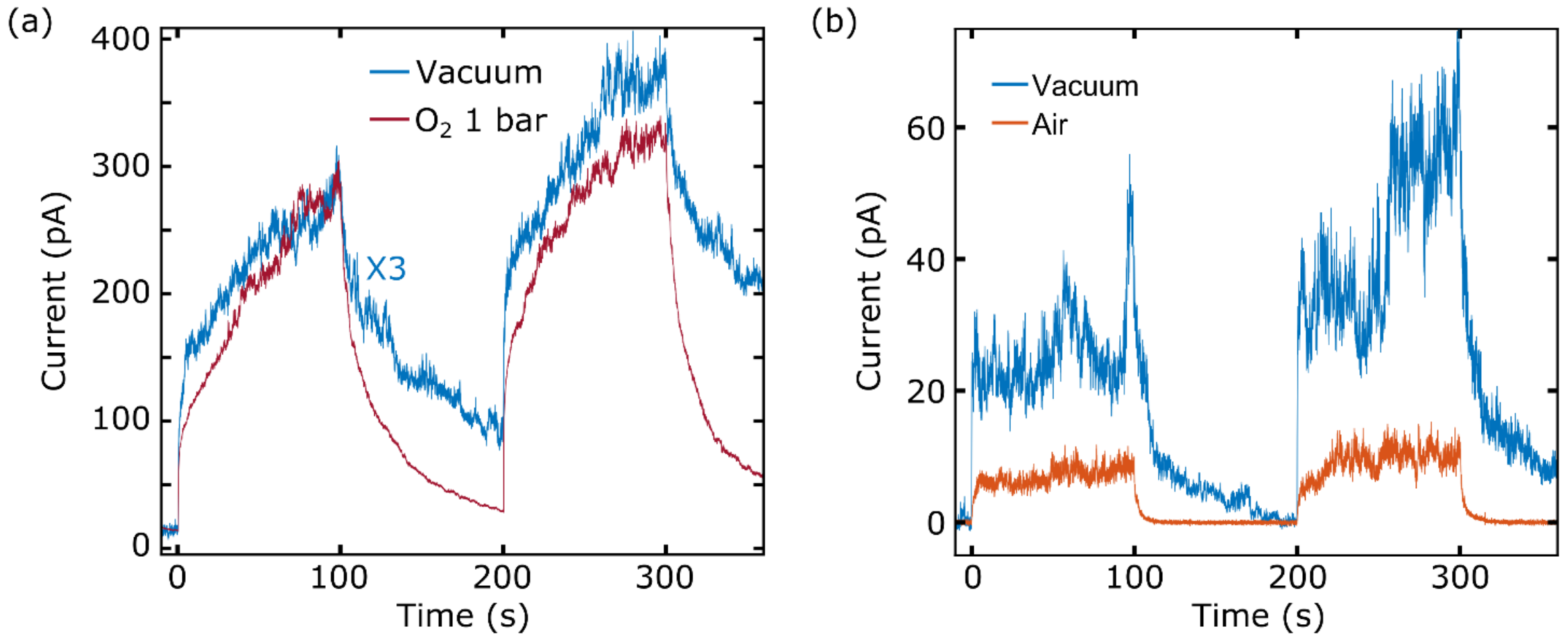

**Fig. S9 (a)** Photoresponse of a device as a function of time in vacuum ($10^{-5}$ mbar) and in oxygen (1 bar). The fall time is considerably slower in vacuum (>100 s) than in oxygen (~90 s). **(b)** Photoresponse of another device as a function of time in vacuum ($10^{-5}$ mbar) and in air. The fall time is considerably slower in vacuum (~80 s) than in air (~1 s).

## Supporting Information references